# Chaotic synchronization in coupled spatially extended beam-plasma systems

Roman A. Filatov [a] Alexander E. Hramov [a],

Alexey A. Koronovskii [a],

[a]*Faculty of Nonlinear Processes, Saratov State University, Astrakhanskaya, 83, Saratov, 410012, Russia*

**Abstract**

The appearance of the chaotic synchronization regimes has been discovered for the coupled spatially extended beam-plasma Pierce systems. The coupling was introduced only on the right bound of each subsystem. It has been shown that with coupling increase the spatially extended beam-plasma systems show the transition from asynchronous behavior to the phase synchronization and then to the complete synchronization regime. For the consideration of the chaotic synchronization we used the concept of time scale synchronization described in work [Chaos, **14**, 3 (2004) 603–610] and based on the introduction of the continuous set of phases of chaotic signal. In case of unidirectional coupling the generalized synchronization regime has been observed in the spatially extended beam-plasma systems. The generalized synchronization appearance mechanism has been analyzed by means of the offered modified system approach [Phys. Rev. E, **71**, 6 (2005) 067201].

# Introduction

The analysis of the chaotic synchronization phenomenon in various nature systems becomes an area of active research of nonlinear science [1–4]. The chaotic synchronization regime has been observed in a whole series of coupled physical, biological, physiological, chemical and other systems [1,4–7]. At present, several different types of chaotic synchronization are known such as generalized synchronization [8], phase synchronization [4,9], lag synchronization [10], intermittent lag [11] and intermittent generalized [12] synchronization behaviour, noise-induced synchronization [13], complete synchronization [14] and time scale synchronization [15,16], which generalizes the above-listed types of chaotic synchronization [15,17,18]. The various methods of analysis should be used to detect the different synchronization regimes [9,10,15,16,19–21].

The presence of the generalized synchronization [8] means that there is a functional relation $\mathbf{x}_2(t) = \mathbf{F}[\mathbf{x}_1(t)]$ between the response $\mathbf{x}_2(t)$ and drive $\mathbf{x}_1(t)$ chaotic systems states after the transient is finished. The phase synchronization [4, 9, 21] means that phase locking of chaotic signals occurs, while this signal amplitudes may remain uncoupled and look chaotic. The regime of coupled oscillations called the lag-synchronization regime, if the dynamics of each subsystem occurs with some shift in time $\tau$: $\mathbf{x}_1(t) \approx \mathbf{x}_2(t-\tau)$. Finally, complete synchronization [14] supposes identical dynamics of chaotic oscillators

*Email addresses:* `filatov_ra@mail.ru` (Roman A. Filatov), `aeh@nonlin.sgu.ru` (Alexander E. Hramov), `alkor@nonlin.sgu.ru` (Alexey A. Koronovskii).

: $\mathbf{x}_1(t) \approx \mathbf{x}_2(t)$. In our works [15, 17] it was shown that generalized, phase, lag- and complete synchronization are closely related with each other and may be considered as the particular cases of the same type of synchronized oscillations which was called time scale synchronization. The character of the synchronized regime (phase, generalized, lag- or complete synchronization) is determined by the quantity of synchronized time scales, which are introduced by the continuous wavelet transform [22–24]. Since the time scale $s$ is bound up with frequency, synchronization of chaotic oscillations is connected with appearance of the phase coupling between spectral components $\omega$ of Fourier–spectra $S(\omega)$ [16].

Most research dealing with chaotic synchronization were realized for the systems with low number of degrees of freedom [2–4, 14] and for the sample models of spatially extended systems (chains and networks of coupled nonlinear chaotic oscillators [3, 25–27], coupled Ginsburg-Landau [3, 28, 29] and Kuramoto-Sivashinsky [30] equations and others). The study of chaotic synchronization in the spatially extended systems was experimentally an theoretically carried out for the nonlinear optical [31–33] and chemical [34] systems and for the low-frequency oscillations in plasma discharge tubes [35,36]. It was shown that while introducing the unidirectional or symmetric coupling in the distributed systems they demonstrate the various types of chaotic synchronization, namely the complete, lag and generalized synchronization. However, the chaotic synchronization in the locally coupled microwave beam-plasma systems is not examined in detail by now. Besides, the transition between various synchronization types are not well investigated and obvious analogies between chaotic synchronization in the spatially extended and low-dimensional systems are not formulated. At the same time research of chaotic synchroniza-tion regimes in spatially extended beam-plasma microwave systems seems to be important because of its applications dealing with data transmission [37–40] and control of chaotic oscillations in microwave electronics systems (see, for example, [41]).

The present paper deals with research of chaotic synchronization in coupled beam-plasma microwave systems with overcritical current — Pierce diode fluid models [42–44], which seems to be interesting as an important model of beam-plasma systems showing various types of chaotic behavior [43–48].

The paper is organized as follows. In Section 1 the fluid model of Pierce diode is briefly discussed. In Section 2 the complete synchronization and time scale synchronization of mutual coupled beam-plasma systems are described. In Section 3 we discuss the generalized synchronization regime in unidirectionally coupled spatially extended beam-plasma systems and describe the method of modified system [49] applied to the discussed model. In conclusion, we summarise the main results discussed in our paper.

## 1 General formalism

Pierce diode [42–44] is one of the simple spatially extended beam-plasma systems demonstrating chaotic dynamics [43–46, 48, 50]. It consists of two infinite parallel plains pierced by a mono-energetic electron beam (Fig. 1). Grids are grounded and the distance between them is $L$. Charge density $\rho_0$ and electron velocity $v_0$ are maintained constant at the system input. The region between two plains is evenly filled by neutralizing stationary ions, whose density $|\rho_i|$ is equal to the non-perturbed beam electron density $|\rho_0|$.

The only one dimensionless control parameter of this system is Pierce parameter

$$\alpha = \omega_p L / v_0$$

, where $\omega_p$ is the electron beam plasma frequency, $v_0$ is the non-perturbed electron velocity, $L$ is the distance between diode plains. This distributed model, though rather simple, demonstrates many features of the electron beam dynamics in different real electron devices such as virtual cathode oscillators (vircators) [51, 52]. With $\alpha > \pi$, the so-called Pierce instability develops in the system and the virtual cathode is formed in the electron beam [42, 44]. At the same time in a narrow range of Pierce parameter values near $\alpha \sim 3\pi$ the increase of the instability is suppressed by the non-linearity and in the electron beam the regime without reflection takes place [44, 52]. In this case the system may be described by fluid equations [43, 44, 52]. It is known [43–46, 48] that in this regime various types of beam-plasma chaotic oscillations occur.

Dynamics of two coupled Pierce diodes in fluid electronic approximation is described by the self-congruent system of dimensionless Poisson, continuity and motion equations:

$$\frac{\partial^2 \varphi_{1,2}}{\partial x^2} = \alpha_{1,2}^2 (\rho_{1,2} - 1), \tag{1}$$

$$\frac{\partial \rho_{1,2}}{\partial t} = -v_{1,2} \frac{\partial \rho_{1,2}}{\partial x} - \rho_{1,2} \frac{\partial v_{1,2}}{\partial x}, \tag{2}$$

$$\frac{\partial v_{1,2}}{\partial t} = -v_{1,2} \frac{\partial v_{1,2}}{\partial x_{1,2}} - \frac{\partial \varphi_{1,2}}{\partial x}, \tag{3}$$

with boundary conditions:

$$v_{1,2}(0,t) = 1, \quad \rho_{1,2}(0,t) = 1, \quad \varphi_{1,2}(0,t) = 0, \tag{4}$$

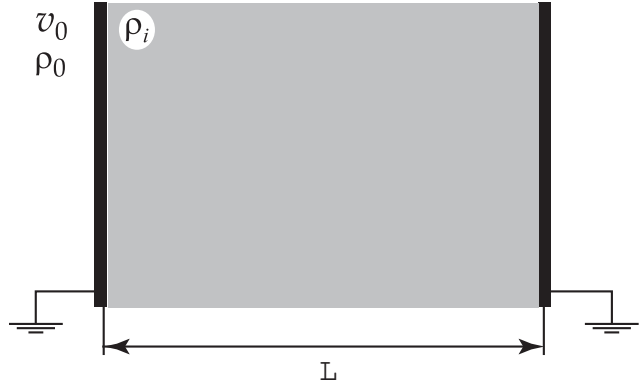

Fig. 1. Schematic diagram of Pierce diode

where indexes "1" and "2" relate to the first and the second coupled beam-plasma systems respectively.

In equations (1)—(3) dimensionless variables of potential $\varphi$, density $\rho$ and velocity $v$ of electron beam, spatial coordinate $x$ ($0 \leq x \leq 1$) and time $t$ are connected with the corresponding dimensional variables

$$\varphi' = (v_0^2/\eta)\varphi, \qquad \rho' = \rho_0\rho,$$

$$v' = v_0 v, \qquad x' = Lx, \qquad t' = (L/v_0)t, \tag{5}$$

where symbols with prime correspond to dimensional variables, $\eta$ is the specific electron charge, $v_0$ and $\rho_0$ is the non-perturbed velocity and density of an electron beam.

In the present paper we study both mutual and unidirectional coupling between Pierce diodes. Mutual coupling between systems is realized by the modulation of dimensionless potential value on the right bounds of both systems:

$$\varphi_{1,2}(x = 1.0, t) = \varepsilon \left\{ \rho_{2,1}(x = 1.0, t) - \rho_{1,2}(x = 1.0, t) \right\}. \tag{6}$$

In case of unidirectional coupling the boundary conditions on potential are defined in the following form:

$$\begin{cases} \varphi_1(1,t) = 0 \\ \varphi_2(1,t) = \varepsilon(\rho_1(x=1,t) - \rho_2(x=1,t)), \end{cases} \quad (7)$$

where the first "1" (drive) system is in the autonomous oscillation regime effecting upon the second "2" (response) system.

Value $\varepsilon$ is the coupling strength between systems in the time depended boundary conditions, which describe mutual (6) and unidirectional (7) coupling, and values $\rho_{1,2}(x=1.0,t)$ present oscillations of non-dimensional density of spatial charge, which are registered at the output of the both systems. It may be carried out experimentally by putting a cutoff of a spiral slow wave system at the output of diode space [53], which register oscillations of electron beam charge density.

System of partial differential equations (1)—(3) has been integrated numerically using finite difference approximation. Numerical solutions for continuity (2) and motion (3) equations have been found by means of explicit scheme with differences against flow, and Poisson equation (1) has been integrated using error vector propagation method [54]. The time and space integration steps have been selected as $\Delta x = 0.005$ and $\Delta t = 0.003$.

It is well known [43–46, 48], that in the autonomous fluid model the transition to chaotic spatiotemporal dynamics of spatial charge is observed via cascade of period doubling bifurcations with $\alpha$ parameter decrease in the range $\alpha \in (2.85\pi, 2.87\pi)$. This is shown in Fig. 2. where space charge density oscillation bifurcation diagram at the point of the interaction space, $x = 0.2$, and dependence of the largest Lyapunov exponent value on Pierce parameter are shown. Value of the highest Lyapunov exponent $\lambda$ has been calculated using Benetdin algorithm adapted for an analysis of distributed system. It is evident from Fig. 2 that with Pierce parameter $\alpha$ decrease the oscillation complexity increases on average.

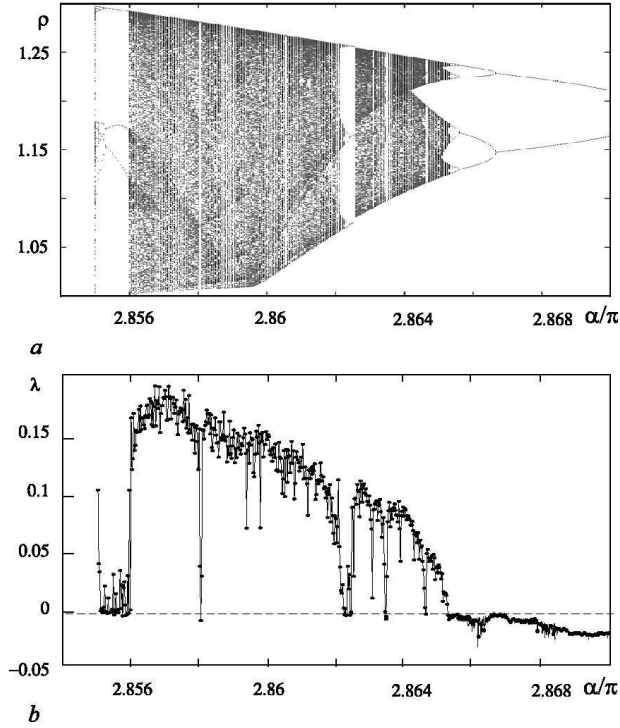

Fig. 2. Bifurcation diagram ($a$) and dependence of the largest Lyapunov exponent ($b$) of autonomous Pierce diode oscillations on $\alpha$ parameter

## 2 Complete and time scale synchronization in mutual coupled distributed chaotic systems

Let us consider the dynamics of the coupled Pierce diode model (1)—(3) with mutual coupling (6) for the fixed Pierce parameter value of the first system $\alpha_1 = 2.861\pi$ and changing the control parameter $\alpha_2$ of the second system in the range $\alpha \in (2.85\pi, 2.87\pi)$. The chaotic dynamics is observed in this range of parameters.

Numerical simulation showed that time scale synchronization may be observed for the small detuning between spatially extended chaotic systems $\Delta\alpha = \alpha_1 - \alpha_2$. The appearance of the time scale synchronization regime is determined by the analysis of phase difference dynamics on the various time scales $s$. A continuous set of phases $\phi_s(t)$ of chaotic time series was introduced with the help of the continuous wavelet transform [16–18]. Chaotic oscillations of spatial charge density $\rho_{1,2}(t)$ at the point $x = 0.2$ of interaction space were examined as an analyzed time series.

The main idea of the time scale synchronization [15–18] is in the following. Let us consider a continuous wavelet transform of time series $\xi(t)$

$$W(s, t_0) = \int_{-\infty}^{+\infty} \xi(t) \psi^*_{s,t_0}(t)\, dt, \tag{8}$$

where

$$\psi_{s,t_0}(t) = \frac{1}{\sqrt{s}} \psi_0 \left( \frac{t - t_0}{s} \right)$$

is the wavelet-function related to the mother-wavelet $\psi_0(t)$. The time scale $s$ corresponds to the width of the wavelet function $\psi_{s,t_0}(t)$, $t_0$ is the shift of wavelet along the time axis ("*" denotes complex conjugation) [22, 24]. It should be noted that the time scale is usually used instead of the frequency of the Fourier transform and can be considered as the quantity inversed to it.

The Morlet wavelet [55]

$$\psi_0(\eta) = (1/\sqrt[4]{\pi}) \exp(j\omega_0 \eta) \exp\left(-\eta^2/2\right)$$

has been used as a mother-wavelet function. The choice of parameter value

$\omega_0 = 2\pi$ provides the relation $s = 1/f$ between the time scale $s$ of wavelet transform and frequency $f$ of Fourier transform.

The wavelet surface

$$W(s, t_0) = |W(s, t_0)| \exp[j\phi_s(t_0)] \qquad (9)$$

describes the system dynamics on every time scale $s$ at the moment of time $t_0$. The value of $|W(s, t_0)|$ indicates the presence and intensity of the time scale $s$ at the moment of time $t_0$. It is also possible to consider the quantity

$$\langle E(s) \rangle = \int |W(s, t_0)|^2 \, dt_0. \qquad (10)$$

which is the distribution of integral energy over time scales. At the same time, the phase $\phi_s(t) = \arg W(s, t)$ is naturally introduced for every time scale $s$. In other words, it is possible to describe the behavior of each time scale $s$ by means of its own phase $\phi_s(t)$.

If there is a range of time scales $[s_m; s_b]$, for which the phase locking condition

$$|\phi_{s1}(t) - \phi_{s2}(t)| < \text{const} \qquad (11)$$

is satisfied and the part of the wavelet spectrum energy (10) in this range $s \in [s_m; s_b]$ is not equal to zero

$$E_{snhr} = \int_{s_m}^{s_b} \langle E(s) \rangle \, ds > 0, \qquad (12)$$

then we assert that time scale synchronization between oscillators takes place [15, 17]. In the condition (11) $\phi_{s1,2}(t)$ are continuous phases of the first and the second systems which correspond to the synchronized time scales $s$.

Introducing a continuous set of time scales $s$ and the instantaneous phases associated with them, as well as separation of the range of synchronous time scales $\Delta s = s_2 - s_1$ allows us to inject the quantitative performance of a measure of chaotic synchronization of coupled systems. This measure $\gamma$ can be defined as a part of wavelet spectrum energy falling on the synchronized time scales:

$$\gamma = \int_{s_m}^{s_b} \langle E(s) \rangle \, ds \bigg/ \int_0^\infty \langle E(s) \rangle \, ds, \qquad (13)$$

where $\langle E(s) \rangle$ is the distribution of integral energy over time scales, which is determined by relation (10). This measure $\gamma$ is equal to zero for non-synchronized oscillations, $\gamma \neq 0$ means that in the coupled systems the conditions of time-scale synchronization (11) and (12) are implemented. The value $\gamma \approx 1$ shows, that oscillations in each of subsystems are identical or close to each other. Such regime is called the complete synchronization regime. Growth of $\gamma$ value from 0 up to 1 testifies the increase of a part of the wavelet spectrum energy, falling on the synchronous time scales $s$.

Let us revert to examination of mutual oscillations in the system of Pierce diode fluid models.

The behavior of coupled beam-plasma systems is illustrated by Fig. 3$a$, plotted for $\alpha_1 = 2.861\pi$ and $\alpha_2 = 2.860\pi$, on which the change of the synchronous time scales range $s_m$ and $s_b$ is shown during an increase of the coupling parameter $\varepsilon$. From Fig. 3$a$ one can see, that for $\varepsilon > 0.0007$ the synchronous time scales exist. This regime, as it was discussed above, corresponds to the time-scale synchronization. The range of synchronous scales increases while increasing coupling parameter $\varepsilon$. For $\varepsilon \approx 0.08 \div 0.1$ the system's dynamics becomes

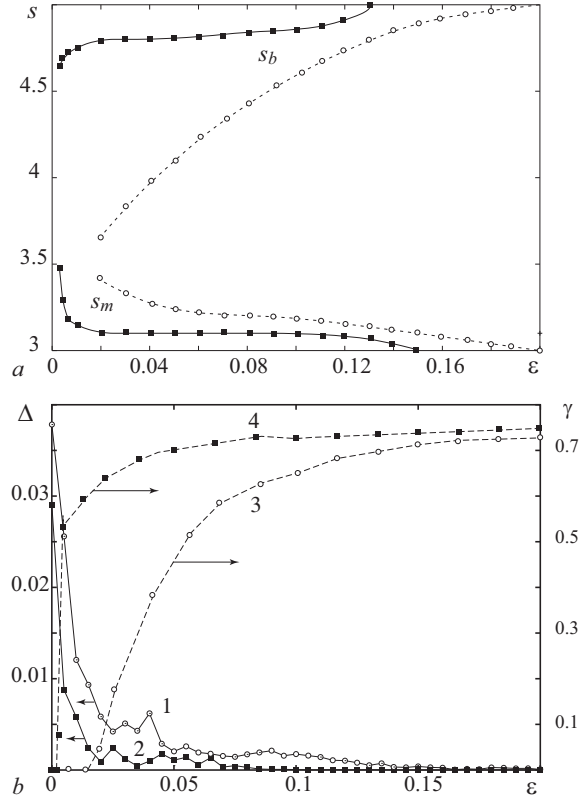

Fig. 3. Dependence of the lowest $s_m$ and the highest $s_b$ bounds of synchronous time scales area (a) and dependencies of the identity measure $\Delta$ of chaotic spatiotemporal oscillations (curves 1,2) and value $\gamma$ of conditional energy (curves 3,4), falling on the synchronous time scales (b), on the value of coupling parameter $\varepsilon$ for the small ($\alpha_1/\pi = 2.860$, $\alpha_2/\pi = 2.861$ (■)) and large ($\alpha_1/\pi = 2.860$, $\alpha_2/\pi = 2.858$ (○)) parameter detuning

synchronous virtually on the whole range of time scales – the regime similar to lag synchronization appears in the coupled beam-plasma systems and the value of time shift between the states of the systems $\tau \approx 0.07$. Further, the time shift decreases with the coupling parameter $\varepsilon$ increase and the coupled system tends to demonstrate the complete chaotic synchronization regime, which is characterized by close to identical dynamics of each of the coupled systems ($\tau \approx 0$).

When greater detuning of coupled beam-plasma system parameters $\Delta\alpha$ occurs, the spectral distribution of oscillations in electron beam becomes essentially more complicated and the appearance of time scale synchronization is observed for the larger values of coupling strength. In Fig. 3a for the case of $\alpha_1 = 2.860\pi$ and $\alpha_2 = 2.858\pi$ the corresponding boundaries $[s_m, s_b]$ of the synchronous time scales area are shown. Similar to the described above, with $\varepsilon$ parameter increase the range of synchronous time scales appears and the system tends to show the complete synchronization regime. However, it occurs under essentially greater values of coupling parameters rather than it was observed in case of small detuning. To analyze the degree of vicinity the dependence of spatiotemporal chaotic oscillations identity measure $\Delta$

$$\Delta = \langle |\rho_1(x,t) - \rho_2(x,t)| + |v_1(x,t) - v_2(x,t)| +$$

$$+ |\phi_1(x,t) - \phi_2(x,t)| \rangle, \qquad (14)$$

on the coupling parameter has been calculated for each of the systems. In equation (14) the symbol $\langle \ldots \rangle$ means the averaging by the time and the space. The results are shown in Fig. 3b (curves 1,2), from which one can see that function $\Delta(\varepsilon)$ decreases quickly tending to zero with the coupling parameter increase. From Fig. 3b (○) one can see that the value $\Delta$, for the large parameter detuning remains unequal to zero (although it becomes sufficiently small for $\varepsilon > 0.17$) apart from the case of small detuning (Fig. 3b (■)). Oscillation regimes, for which $\Delta(\varepsilon) \approx 0$, should be considered as the complete chaotic synchronization regime.

As it was noted above, the synchronization measure $\gamma$ (13) is an important energetic characteristic of the synchronous behavior of coupled spatially ex-tended chaotic systems. In Fig. 3b (curves 3,4) the dependencies $\gamma(\varepsilon)$ are shown for the sets of control parameters $\alpha_{1,2}$ mentioned above. It is easy to see that the increment of a part of chaotic spatiotemporal oscillation energy falling on the synchronous time scales takes place with the coupling parameter increase. It corresponds to the convergence of oscillations in each of spatially extended systems and, as a result, to the appearance (for the greater values of coupling parameter) of the complete synchronization regime.

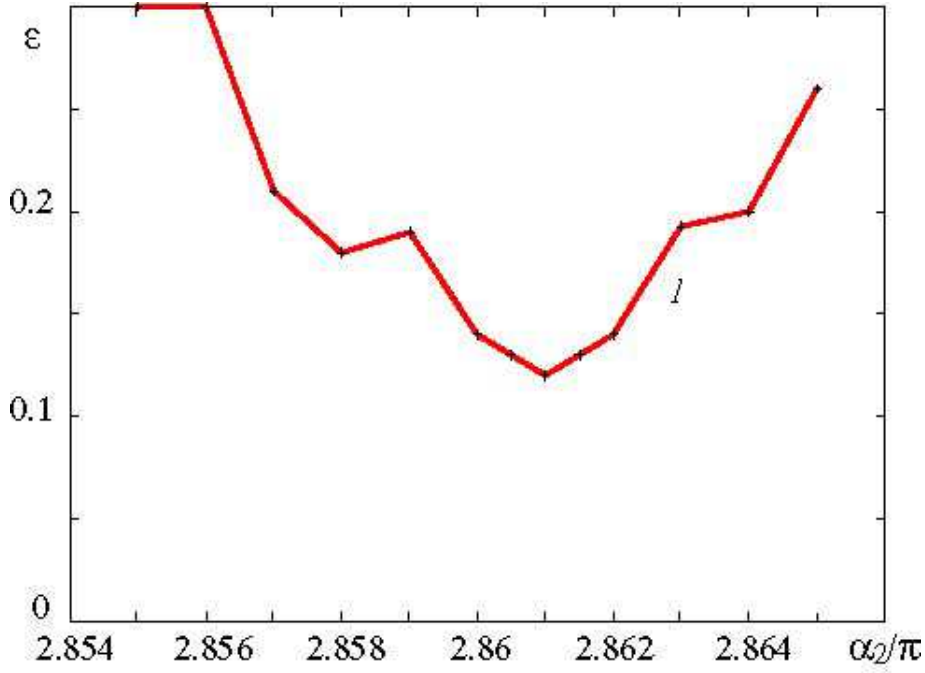

Fig. 4. Control parameter plane $(\alpha_2/\pi, \varepsilon)$ for the fixed value of $\alpha_1 = 2.861\pi$. Boundary of the complete synchronization regime is shown

In Fig. 4 the boundary of complete synchronization area plotted for fixed value of Pierce parameter $\alpha_1 = 2.861\pi$ is shown on the control parameter plane $(\alpha_2, \varepsilon)$. From the figure one can see that the complete synchronization regime takes place with the coupling parameter increase for all detunings of control parameters $\alpha_{1,2}$ of either Pierce diode. It is fair both for the weak chaotic oscillations and small parameter detuning and for developed chaos even for great detuning between control parameters of each subsystem. The minimal values of coupling parameter $\varepsilon$, for which the complete synchronization in spatially extended beam-plasma systems is observed, have been detected for the small detuning of coupled subsystems.

## 3 Generalized synchronization in unidirectionaly coupled Pierce diodes

Let us examine the generalized synchronization phenomenon in the considered coupled spatially extended systems. The generalized synchronization has been introduced only for the unidirectionaly coupled systems, therefore let us consider the unidirectional coupling (7) between spatially extended Pierce beam-plasma systems being in the chaotic oscillations regime.

We used auxiliary system approach [20] and the highest conditional Lyapunov exponent calculating [19, 56] for the generalized synchronization regime diagnostic . The main idea of the auxiliary system approach is that along with the response system $\mathbf{x}_r(t)$ another (auxiliary) system being identical to it, $\mathbf{x}_a(t)$, is examined. Initial conditions for the auxiliary system $\mathbf{x}_a(t_0)$ have to be chosen different from the response system initial state $\mathbf{x}_r(t_0)$ but they should belong to the same chaotic attractor. In the case of lack of the generalized synchronization regime between interacting systems the state vectors of response $\mathbf{x}_r(t)$ and auxiliary $\mathbf{x}_a(t)$ systems are different. And in the case of the generalized synchronization regime presence, the states of the response and auxiliary systems must be identical $\mathbf{x}_r(t) \equiv \mathbf{x}_a(t)$ after the transient in order to fulfill relations: $\mathbf{x}_r(t) = \mathbf{F}[\mathbf{x}_d(t)]$ and, correspondingly, $\mathbf{x}_a(t) = \mathbf{F}[\mathbf{x}_d(t)]$. Thus, the equivalence of the response and auxiliary systems states after the transient (which may be sufficiently long [12]) is a criterion of the generalized synchronization regime presence between the drive and response oscillators.

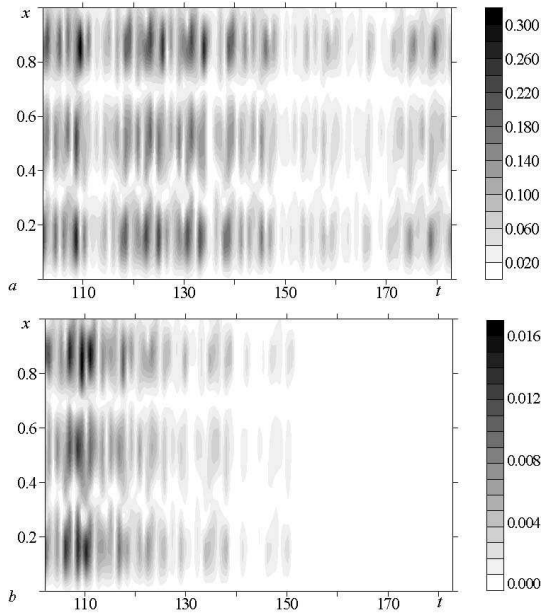

fig. 5. Spatiotemporal evolution of the state difference $|\rho_2(x,t) - \rho_{2a}(x,t)|$ (values of spatial charge density) between the response and drive systems. (*a*) $\varepsilon = 0.05$, the lack of the generalized synchronization regime, (*b*) $\varepsilon = 0.2$, the chaotic generalized synchronization regime: after transient is finished in the response and drive systems the same spatiotemporal states appear

Analysis of the generalized synchronization regime also may be realized by means of the conditional Lyapunov exponent calculating [57]. In this case Lyapunov exponents are calculated for the non-autonomous response system, and since the behavior of this system depends on the drive system behavior these Lyapunov exponents are called conditional. Negativity of the largest conditional Lyapunov exponent $\lambda$ is a criterion of the generalized synchronization presence in unidirectionally coupled dynamical systems [19, 56].

While using auxiliary system approach the system of equations (1)—(3) has been solved for the auxiliary system Pierce parameter value being equal to the response system control parameter, but with different initial conditions. It is useful for the generalized synchronization regime diagnostics to plot the oscillation difference $|\rho_2(x,t) - \rho_a(x,t)|$ between the response ($\rho_2(x,t)$) and auxiliary ($\rho_a(x,t)$) system over all interaction space. Results are presented in Fig. 5, from which one can see that the charge density oscillations in the response and auxiliary systems remain different over all interaction space for smaller values of coupling parameter (Fig. 5$a$), and for the sufficiently large values of $\varepsilon$ oscillations in the response and auxiliary systems become identical (Fig. 5$b$), i.e. the generalized synchronization regime appears in the coupled system.

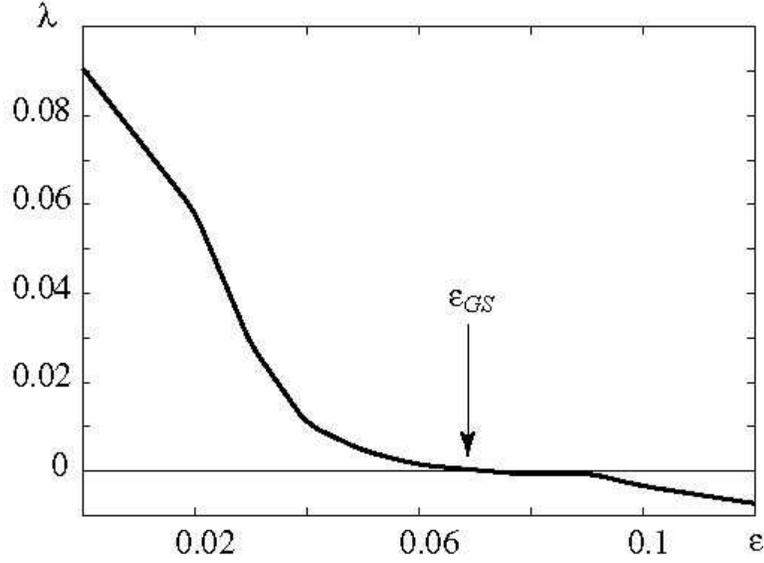

Fig. 6. Dependence of the highest conditional Lyapunov exponent on the coupling parameter $\varepsilon$ for the values of Pierce parameter of the drive and response systems: $\alpha_1 = 2.858\pi$ and $\alpha_2 = 2.862\pi$. The value of the coupling parameter for which the generalized synchronization regime appears in the system (the largest conditional Lyapunov exponent becomes negative) is marked by an arrow

Results obtained by using auxiliary system approach has been confirmed by means of the highest conditional Lyapunov exponent $\lambda$ calculating. Numerical

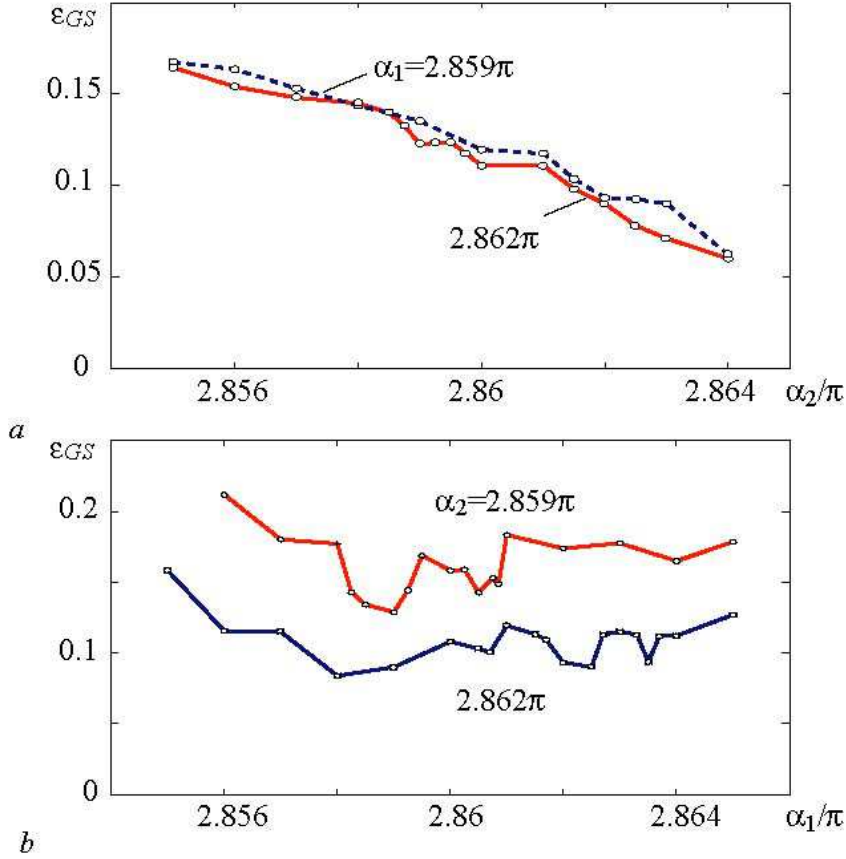

Fig. 7. Dependence of generalized synchronization threshold $\varepsilon_G S$ on the control parameter of the drive $\alpha_2$ (fig. *a*) and response $\alpha_1$ (fig. *b*) distributed systems

calculation of the highest conditional Lyapunov exponent, as above, has been realized by Benetdin's algorithm adapted for distributed system analysis. In Fig. 6 a typical dependence of the highest conditional Lyapunov exponent on the value of the coupling parameter is shown for the values of parameters $\alpha_1 = 2.858\pi$ and $\alpha_2 = 2.862\pi$. One can see that for certain value of the coupling parameter $\varepsilon = \varepsilon_{GS}$ (marked by an arrow in Fig. 6) the generalized synchronization regime appears in the spatially extended beam-plasma systems. Thus, with the coupling parameter increase the appearance of the generalized synchronization is observed in the spatially extended beam plasma system.

When fixing the first system control parameter and changing parameter of the second system one can plot the dependence of the generalized synchronization appearance threshold $\varepsilon_{GS}$ on the detuning between systems. In Fig. 7$a$ such dependence is plotted for the fixed value of drive system control parameter; one can see that with response system control parameter increase (in other words with transition to the area of simpler oscillations Fig. 2) the threshold of the generalized synchronization arising decreases. In Fig. 7$b$ the value of the threshold of the generalized synchronization regime $\varepsilon_{GS}$ has been plotted for the fixed value of the response system control parameter $\alpha_2$ when changing parameter $\alpha_1$ of the drive system. One can see that for the small detuning between the systems the value of the generalized synchronization threshold depends weakly on the drive system parameter.

In our works [29, 49] the various ways of the generalized synchronization arising in the coupled less-dimensional systems and coupled spatially extended Ginzburg-Landau equations were discussed using the modified system approach. The main idea of this method is the substitution of two unidirectional coupled oscillator systems by the modified response system being under external influence of the drive system signal. This method was very helpful and effective for the analysis of the generalized synchronization arising mechanisms connected with presence of additional dissipation in the non-autonomous chaotic system in the generalized synchronization regime [49].

Let us apply this method to the analysis of the generalized synchronization arising mechanism in the coupled spatially extended beam-plasma systems. It is necessary to examine the dynamics of the autonomous modified response system taking into account the introducing of the additional dissipative term in the same way as it has been done in [49]. In our case an autonomous modified distributed system is described by the system of equations (1)—(3) with the following boundary conditions for the potential

$$\begin{cases} \varphi_{2m}(0,t) = 0 \\ \varphi_{2m}(1,t) = -\varepsilon\rho_{2m}(1,t)). \end{cases} \quad (15)$$

In this form the modified distributed system (1)—(3) may be considered as a Pierce diode fluid model with a connected feedback loop. Similar system has been sufficiently investigated in our work [48, 58, 59], here Pierce diode fluid model with external feedback has been examined. It has been shown for this model, that with the feedback coefficient (in the investigated case $\varepsilon$ parameter) increase the transition from chaotic dynamic to periodical oscillations is observed through the cascade of the period doubling bifurcations.

As it has been shown in work [49], the existence of the regular oscillations or steady states in the modified system is an essential condition for the generalized synchronization regime arising. Therefore, taking into account the results of work [48], we can assert that, the mechanism of generalized synchronization regime arising [49] is realized, but in the investigated case it is determined by a reconstruction of oscillation regimes in modified system by means of connection of the feedback of some sort rather than introducing of additional dissipation.

It is illustrated by Fig. 8$a$, on which the bifurcation diagram of Pierce diode with feedback is plotted while changing the control parameter $\varepsilon$. From the figure one can see that, oscillations in the modified system become periodical with coupling parameter increase and after that the arising of the steady state is observed. However, the value of the coupling parameter, for which the arising of the periodical oscillations in system is observed, is appreciably smaller than liminal value of $\varepsilon_{GS}$, for which the generalized synchronization regime arises. As it was discussed in work [49] it is necessary to consider the modified system under the external influence to find the generalized synchronization threshold. In this case, the external influence results in an increase of the parameter $\varepsilon$ value, for which the periodical oscillations is observed. Such behavior may be illustrated by the examination of the modified system dynamics under the external periodical influence. In the simplest case, this influence may be assigned harmonic, whose frequency and amplitude must correspond to the main base frequency of the power spectrum of the drive chaotic system.

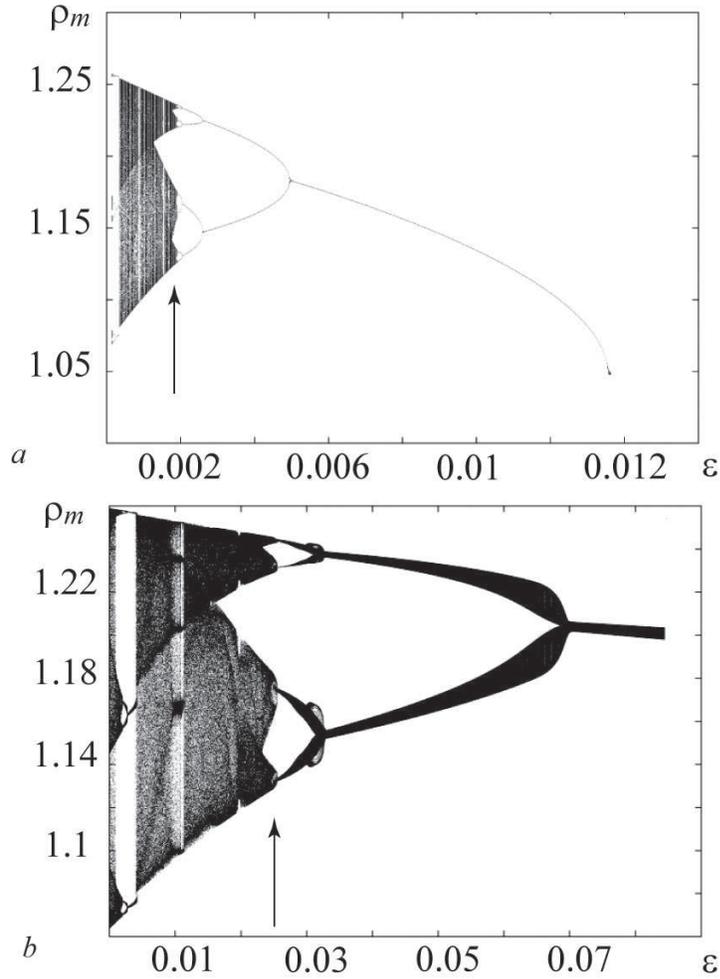

Fig. 8. Bifurcation diagrams of the spatial charge density oscillations in the modified system in case of autonomous dynamics (*a*) and under the external harmonic signal (*b*). Pierce parameter of modified system is $\alpha_2 = 2.862\pi$

In this case we change the boundary conditions (15) for modified system (1)—(3) adding a harmonic signal:

$$\begin{cases} \varphi_m(0,t) = 0 \\ \varphi_m(1,t) = -\varepsilon\rho_r(1,t)) + \varepsilon A\cos(2\pi f_0 t), \end{cases} \quad (16)$$

where $A = 0.78$ and $f_0 = 1$ have been chosen to simulate the main peak in the power spectrum of the drive system.

In Fig. 8$b$ Pierce diode oscillation bifurcation diagram is shown for the case of external influence. One can see that, the bifurcation points of the modified system under the external influence are shifted towards lager values of $\varepsilon$ in comparison with the autonomous case (Fig. 8$a$). At the same time, two-frequency oscillations with incommensurable time scales (quasi-periodical regular oscillations) are observed in the system with the large values of feedback coefficient $\varepsilon$. This may be easily seen from the bifurcation diagram. Therefore, the value of parameter for which the generalized synchronization appears accepts lager values that in autonomous case.

**Conclusions**

Thus, the possibility of arising of the various types of chaotic synchronization (complete synchronization and time-scale synchronization) in mutually coupled beam-plasma systems with overcritical current (coupled Pierce diode fluid models) has been reported for the first time in the paper. It is impor-tant to note that the coupling was introduced only in one point of interaction space. A new approach to chaotic synchronization — time-scale synchronization [15, 17] has been used for the analysis of chaotic synchronization. It is significant that the possibility of the complete synchronization regime appearance of the chaotic spatiotemporal beam-plasma oscillation makes possible the application of such self-oscillating media for the data transmission systems of the microwave range.

In case of unidirectional coupling the transition from asynchronous behavior to the generalized synchronization regime has been observed with the coupling parameter increase. For the fixed control parameter (Pierce parameter) of the drive system the generalized synchronization threshold decreases with the response system Pierce parameter increase. For the fixed value of the response system Pierce parameter the value of coupling parameter corresponding to the generalized synchronization regime arising depends weakly on the changing of the drive system Pierce parameter. Such behavior of the generalized synchronization appearance threshold was explained using the modified system method proposed in work [49] for the analysis of the generalized synchronization in the less-dimensional systems.

This work has been supported by CRDF (grant REC–006) and Russian Foundation for Basic Research (projects 05–02–16286, 06–02–16451). A.E.H. and A.A.K. also thank the Dynasty Foundation.